\documentclass[prl,twocolumn,amsmath,amssymb,floatfix]{revtex4}

\usepackage{times,amsmath}
\usepackage{graphicx}
\usepackage{color,ulem}

\begin{document}

\preprint{APS/123-QED}

\title{DNA translocation through nanopores with salt gradients: The
  role of osmotic flow} \author{Marius M. Hatlo$^\dagger$}
  \email{M.M.Hatlo@uu.nl} \author{Debabrata Panja$^\ddagger$}
  \author{Ren\'e van Roij$^\dagger$} \affiliation{ $^\dagger$Institute
  for Theoretical Physics, Utrecht University, Leuvenlaan 4, 3584 CE
  Utrecht, The Netherlands\\ $^\ddagger$Institute for Theoretical
  Physics, Universiteit van Amsterdam, Science Park 904, Postbus
  94485,  1090 GL Amsterdam, The Netherlands}


\begin{abstract} 
  Recent experiments of translocation of double stranded DNA through
  nanopores [M. Wanunu \textit{et al.}  Nature Nanotech. {\bf 5}, 160
  (2010)] reveal that the DNA capture rate can be significantly
  influenced by a salt gradient across the pore. We show that osmotic
  flow combined with electrophoretic effects can quantitatively
  explain the experimental data on the salt-gradient dependence of the capture rate. 
\end{abstract}
\maketitle
Translocation through solid-state nanopores holds the potential to be
a fast commercial method for macromolecular characterization and
sequencing, such as for long, unlabelled single-stranded and
double-stranded DNA molecules
\citep{Branton_etal_2008,Dekker_2007,Li_etal_2001}. Clearly, high
throughput and time resolution --- effected by enhanced capture rate
as well as translocation times respectively --- is a necessary
precondition for the process to be commercially viable. Although the
capture rate or translocation time can be increased by manipulation of
the temperature, salt concentration, electric field strength and
viscosity \citep{Fologea_etal_2005}, the increase of one is usually
accompanied by a decrease of the other \citep{Fologea_etal_2005}.
Recently however, Wanunu \textit{et al.} \citep{Wanunu_etal_2010}
showed that it is possible to increase the capture rate without
decreasing the translocation time by using a forward salt
concentration gradient across the pore. The large increase in capture
rate as a function of the imposed salt concentration gradient was
qualitatively explained by the increase in the electrophoretic motion
of DNA towards the pore as a function of salt asymmetry: A constant
current of ions is flowing through the pore, creating a long range
electric field which acts as a funnel for the ions and the polymers
towards the pore
\citep{Wanunu_etal_2010,Chou_2009,Grosberg_Rabin_2010}.

These studies \citep{Wanunu_etal_2010,Chou_2009,Grosberg_Rabin_2010},
and some others on similar systems
\cite{Wong_Muthukumar_2007,Ghosal_2007,Luan_Aksimentiev_2008}, have
focused on electroosmotic and/or electrophoretic
phenomena. Electroosmotic phenomena describe flow of liquids with a
net mobile charge under applied electric fields, while electrophoretic effects
relates to the movement of charged polymers in an electric
field. [These terminologies are further discussed in Sec. I of the
Supplementary Online Material]. It is worthwhile to note at this point
that electroosmotic flow in the experiments of Wanunu {\it et al.}
was found to be in the opposite direction of the observed DNA
translocation (see the supplementary material of
Ref. \cite{Wanunu_etal_2010}), implying that electroosmotic flow
cannot explain the observed enhanced capture rate. In fact, in the
presence of an imposed salt concentration gradient across the pore,
there is an additional mechanism at play, namely the capillary osmosis
process
\cite{Ajdari_Bocquet_2006,Derjaguin_etal_1972,Palacci_etal_2010},
i.e., {\it the flow of water from a lower osmotic pressure ({\it cis})
side to a higher osmotic pressure ({\it trans}) side through the
pore.\/} The effects of this osmotic flow on the DNA capture rate has
been missing in the theoretical analysis so far.  In this Letter we
show that the osmotic flow is a key ingredient to understand the
experiments of Wanunu \textit{et al.}  \citep{Wanunu_etal_2010}; their
results can be \textit{quantitatively} explained with osmotic flow and
electrophoretic effects \citep{Derjaguin_etal_1972}. Note also that
the full range of dynamical mechanisms affecting DNA capture in the
experimental setup of Wanunu {\it et al.} are discussed in Sec. I of
the Supplementary Online Material.

The osmotic flow of water is driven by a pressure gradient
anti-parallel to the salt concentration gradient. The reservoirs are
kept at a constant pressure, such that a chemical potential gradient
is present accross the pore for both the ions and water, causing flow
of ions down the salt concentration gradient and water up the salt
concentration gradient. However, a pressure gradient inside the pore
and a correspondig net flow of the liquid (ions plus water) will only
be present if the ions are net depleted from (or net attracted to) the
pore \cite{Anderson_Malone_1974}. In the textbook example of osmosis
through a semi-permeable membrane ions are completely restricted from
entering the pore due to steric repulsion
\citep{Anderson_Malone_1974}. However, also when the restriction is
only partial, e.g. due to wall-ion interactions in the \textit{nm}
vicinity of the pore walls, an osmotic flow develops
\citep{Anderson_Prieve_1991,Ajdari_Bocquet_2006,Derjaguin_etal_1972,Huang_etal_2008b}.
Water and ions confined in a nanopore can behave very differently from
the same bulk system
\cite{Kalcher_etal_2010,Beckstein_etal_2004,Parsegian_1969,Yaroshchuk_2000,Hatlo_etal_2008,Schwierz_etal_2010,Levin_etal_2009}. Both
water and ions will be influenced by the pore walls, leading to
attraction or depletion of ions and/or water. To capture such a
behavior with a simple continuum model we introduce one length scale
describing the interaction of the ions with the pore wall (the
depletion length), and one length scale quantifying the slip of water
flow at the pore wall (the slip length).  To quantitatively describe
the experimental data of Wanunu \textit{et al.}
\cite{Wanunu_etal_2010}, the ion-wall interactions is found to be
repulsive, in agreement with simulations
\cite{Huang_etal_2008b,Schwierz_etal_2010} and theories
\cite{Levin_Mena_2001,Hatlo_etal_2008} of kosmotropic (hydrophilic)
ions near low dielectric surfaces. Also, the flow of water is found to
be in the slip-flow regime, in agreement with experimental studies of
flow at smooth surfaces \cite{Bonaccurso_etal_2003}.  In the
experimental system studied by Wanunu \textit{et al.}
\cite{Wanunu_etal_2010} we find the osmotic flow to provide the
dominant contribution to the enhanced capture rate for weak salt
gradients.  
In the same experiment \cite{Wanunu_etal_2010} electroosmosis was found be a weak effect, reflected in a two/three orders of
magnitude lower capture rate of neutral PEG compared to charged DNA. Also the measured purely Nernstian behavior of the diffusion potential suggests nearly neutral pore walls.

\begin{figure}[t] \begin{center} \resizebox{\columnwidth}{!}{
 \includegraphics[clip]{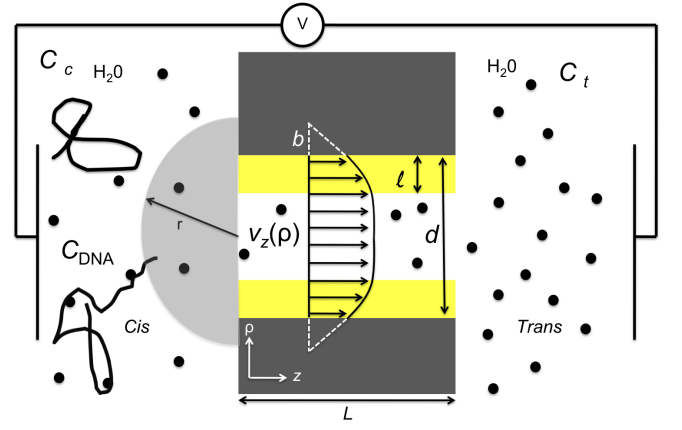}}
    \caption{\label{fig:schematic} Schematic of the pore geometry
      showing a membrane of thickness $L$ connecting two salt
      reservoirs with salt concentrations $C_c$({\it cis}) and $C_t$
      ({\it trans}) by a pore of diameter $d$. In the {\it cis}
      reservoir there is a bulk DNA concentration of $C_{\rm DNA}$,
      and a voltage difference $V$ is applied accross the system. The
      salt ions are depleted within a layer $\ell$ from the pore
      walls. There is a liquid velocity profile $v_z(\rho)$ in the $z$
      direction, which varies with the radial coordinate $\rho$. There
      is slip of the flow at the pore walls described by the slip
      length $b$.}
\end{center}
\end{figure}

 The geometry we study, similar to the experimental setup of
Ref. \cite{Wanunu_etal_2010}, is shown in Fig.
\ref{fig:schematic}. Two reservoirs at constant pressure $P_0$ with
salt concentration $C_t$ ({\it trans} side) and $C_c$ ({\it cis} side)
are separated by an impermeable solid membrane of thickness $L$. A
cylindrical pore of diameter $d$ connects the two reservoirs.  The two
electrolytes are composed of monovalent ions of concentrations
$c_\alpha$, and $\sum_{\alpha=\pm} c_\alpha=C$. The solvent (water) is
modeled as a continuum with dielectric constant $\epsilon = 80$, and
viscosity $\eta$ at temperature $T$. The Debye screening lengths
$\kappa^{-1}_{c/t}$, are defined as $\kappa^2_{c/t}= 4\pi C_{c/t}\beta
e^2/\epsilon$, where $\beta^{-1} = k_B T$, $k_B$ is the Boltzmann
constant and $e$ is the elementary charge. Due to the preference of
ions to be solvated in bulk water, they feel a repulsive potential
$U(\rho)$ from the pore walls \citep{Hatlo_etal_2008,Parsegian_1969},
where $\rho$ is the radial coordinate around the cylindrical axis
inside the pore, measured from the center of the pore. We model such
interactions with a region $\ell$ next to the pore walls depleted of
ions (see Fig.  \ref{fig:schematic}). The part of the pore accessible
to ions is described by the diameter $a=(d-2\ell)$. The polymers (DNA)
are located in the {\it cis} chamber, and the electric field is
applied from the {\it trans} to {\it cis} side, driving DNA (with
negative charge) from the {\it cis} to the {\it trans} reservoir.

The pore is assumed to be neutral, i.e. $\sum_\alpha q_\alpha
c_\alpha(\rho,z) = 0$ \citep{Wanunu_etal_2010}, where $q_\pm = \pm e$,
which is a good approximation for nearly neutral pore walls and low
induced charge within the pore.

For the system studied here, the Reynolds number is very small, and
the flow can be described by the steady Stokes equation combined with
incompressibility of the liquid:
\begin{eqnarray}
\label{eq:NS}
\eta\nabla^2{\bf v}(\rho,z) &=& \nabla P(\rho,z) +\sum_\alpha
c_\alpha(\rho,z)\nabla U(\rho); \\
\label{eq:incomp}
\nabla\cdot {\bf v} &=& 0.
\end{eqnarray}
In steady state the dynamics of the ions are described by the
time-independent Nernst-Planck equations:
\begin{eqnarray}
\label{eq:NP}\nonumber
\nabla\cdot {\bf J_\alpha}& = & -D_\alpha
\Big{(}\nabla^2c_\alpha(\rho,z)\\ &&+\nabla
\cdot(c_\alpha(\rho,z)\beta[\nabla U(\rho)-q_\alpha{\bf E}(z)])\Big{)}
\\\nonumber &&+\nabla \cdot [c_\alpha(\rho,z){\bf v}(\rho,z)]=0.
\end{eqnarray}
This is equivalent to conservation of particle current, with ${\bf
  J_\alpha}$ the current density, $D_\alpha$ the diffusion coefficient
  of ion type $\alpha$ and ${\bf E}(z)$ the local electric field.
Assuming fast equilibration of the concentration and the pressure in
the radial direction ($\hat{\rho}\cdot{\bf J}_\alpha=0$ and
$\hat{\rho}\cdot {\bf v}=0$) we get from Eqs. \eqref{eq:NS} and
\eqref{eq:NP} \citep{Ajdari_Bocquet_2006}
\begin{eqnarray}
c_\alpha(\rho,z) &\approx& \left\{\begin{array}{llll} C_0(z) &  & &
\rho<a/2 \\ 0 &  & &  \rho>a/2.
\end{array}
\right.  \\ \eta\nabla^2v_\rho(\rho) &=& 0,
\end{eqnarray}
which from Eq. \eqref{eq:NS} gives
\begin{equation}
\partial_zP(\rho,z) =  \left\{\begin{array}{llll} 0 & & &  \rho<a/2 \\
-k_BT \partial_z C_{0}(z) & & &   \rho>a/2.
\end{array}
\right. 
\end{equation} 
If we further assume that the ion density changes linearly accross the
pore (which follows from Eq. \eqref{eq:NP} when diffusion dominates
over convection) we get
\begin{equation}
\label{eq:NSz}
\eta \nabla^2 v_z(\rho)= \left\{\begin{array}{llll} 0 & & & \rho<a/2
\\ -k_B T\displaystyle\frac{C_t-C_c}{L} & & & \rho>a/2.
\end{array}
\right.
\end{equation}
Since the pore is in the nanometer regime a continuum treatment of the
fluid dynamics may not be accurate. The Knudsen number for the system
is $Kn = \delta/d \approx 0.1$, where $\delta \approx 3$ \AA$ $  is
the intermolecular spacing for water \cite{Karimi_Li_2005}, indicating
that we are in the slip-flow regime ($0.01\leq Kn\leq 0.1$), such that
\begin{equation}
\label{eq:slip}
v_z(d/2) = b \left.\frac{\partial v_z(\rho)}{\partial
\rho}\right|_{\rho=d/2}, 
\end{equation}
where $b$ is the slip length (see Fig.\ref{fig:schematic}).  The flow
can now be obtained by integrating Eq. (\ref{eq:NSz}) twice making use
of Eq. (\ref{eq:slip}). The resulting area-averaged velocity of the
flow inside the pore is
\begin{equation}
\label{eq:barvo}
\bar{v}_\text{o} = \frac{k_B T(C_t-C_c)\sigma_{\text
  o}}{L}\frac{d^2(1+8b/d)}{32\eta},
\end{equation}
where we have introduced the osmotic reflection coefficient
\begin{equation}
\label{eq:sigmao}
\sigma_\text{o} =
1-\frac{(a/d)^2}{1+8(b/d)}\left(8(b/d)+2-(a/d)^2\right).
\end{equation}
For $a=0$ ($\sigma_\text{o} = 1$), i.e. ions are totally depleted from
the pore, we recover the standard slip-modified Poiseuille flow due to
osmosis through a semipermeable membrane \citep{Thomas_etal_2010}. If
we set the slip length to zero, we recover the result of Anderson and
Malone for leaky membranes\citep{Anderson_Malone_1974}, however with
an effective solute radius $\ell$. With $\ell=0$ (no ionic depletion)
Eq. \eqref{eq:sigmao} gives $\sigma_\text{o}=0$ (no flow), showing
that ion depletion is crucial for the present analysis.  From
Eqs. (\ref{eq:NS}) and (\ref{eq:incomp}) the osmotic flow at a radial
distance $r\gg d$ from the pore can now be approximated as
\begin{equation}
\label{eq:vo}
{\bf v}_{\rm OS}(r) =
-\hat{r}\frac{\bar{v}_\text{o} d^2}{8 r^2},
\end{equation}
where $\hat{r}$ is the radial unit vector, pointing outward from the
pore mouth. 

In a steady state and using conservation of charge current
(Eq. \eqref{eq:NP}), the electric field on the {\it cis} side (for
$|r|\gg d$) can be approximated as \citep{Wanunu_etal_2010}:
\begin{equation}
{\bf E}(r) = \hat{r}\frac{C_{p} a^2 V}{8C_{c} L r^2},
\end{equation}
where $C_{p} = (C_{t}+C_{c})/2$ is the ion concentration inside the
accessible part of the pore, and $V/L$ is the strength of the applied
$E$-field in the pore. The drift of charged polymers in an electric
field is described by electrophoresis \citep{Firnkes_etal_2010}
\begin{equation}
  \label{eq:EP}
{\bf v}_{\rm EP}(r) =\mu{\bf E}(r)=\frac{\phi_{\rm DNA}\epsilon}{4\pi
    \eta} {\bf E}(r),
\end{equation}
where $\phi_{\rm DNA}$ is the surface potential of DNA, and $\mu$ is
the electrophoretic mobility.

To get an estimate of the number of polymers that translocate through
the pore per second, we calculate the flux of DNA generated by the
combination of electrophoretic effects and osmosis.  By
conservation of DNA particle current, we get the capture rate per bulk
DNA concentration
\begin{equation}
\label{eq:RC}
R_c = -2\pi r^2 \Big{[}{\bf v}_{\rm EP}(r)+{\bf v}_{\rm OS}(r)
\Big{]}\cdot \hat{r},
\end{equation}
independent of $r$. Flow towards the pore is antiparallel to $\hat{r}$
(see Fig.  \ref{fig:schematic}), and therefore a negative sign in Eq.
\eqref{eq:RC} appears such that $R_c >0$ for translocation from the
{\it cis} to {\it trans} reservoir. Combining Eqs. \eqref{eq:vo},
\eqref{eq:EP} and \eqref{eq:RC} we find:
\begin{equation}
\label{eq:capture_rate}
R_c(x) = R_c(1)\left[\frac{1+x}{2} +
  \left(1-\frac{1}{x}\right)k\right],
\end{equation}
where $x = C_t/C_c$ and
\begin{eqnarray}
\label{eq:capture_sym}
R_c(1) &=& \frac{a^2\phi_{\rm DNA}V\epsilon}{16 L\eta}\\
\label{eq:k}
k &=& \frac{(\kappa_t d)^2\sigma_\text{o}(1+8b/d)}{32 (a/d)^2(\beta e
\phi_{\rm DNA})(\beta e V)}.
\end{eqnarray}
Note that the result does not depend on DNA length
\citep{Wanunu_etal_2010}, and that $k=0$ (or $x=1$) describes electrophoretic
effects alone.
\begin{figure}[h] \begin{center} \resizebox{\columnwidth}{!}{
      \includegraphics[clip]{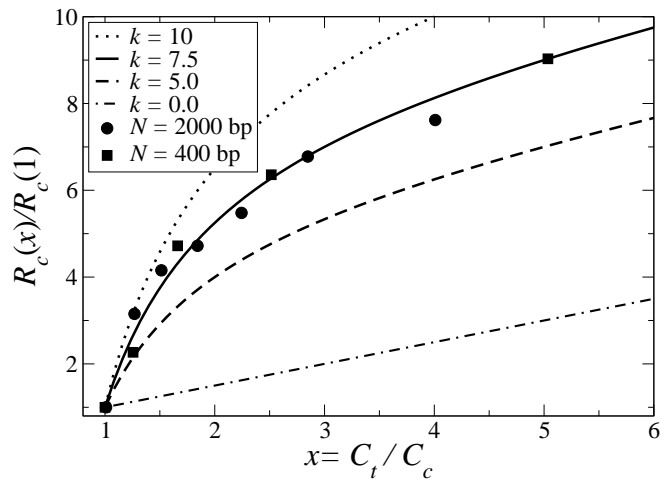}}
    \caption{\label{fig:Rate_enhancement_k}
      Capture rate (Eq. (\ref{eq:capture_rate})) as a function of salt
      asymmetry, for different values of the dimensionless parameter
      $k$ (see Eq. \eqref{eq:capture_rate}). The points are
      experimental mesurements of Ref. \citep{Wanunu_etal_2010}.} 
\end{center}
\end{figure}

In Fig. \ref{fig:Rate_enhancement_k} we plot the predictions of Eq.
(\ref{eq:capture_rate}) as a function of salt asymmetry $x$ for
different values of the dimensionless parameter $k$ (Eq.
(\ref{eq:k})). As the value of $k$ increases the predictions start to
deviate from the capture rate due to electrophoretic effects alone
(straight line, $k=0$). The flow due to osmosis varies inversely with
$x$ (i.e. linear in $C_c$), and will therefore saturate for large $x$,
while the flow due to electrophoretic effects has a linear
dependence on $x$ with slope $1/2$. With $k$ in
Eq. (\ref{eq:capture_rate}) as a free parameter, we find an excellent
fit to the experimental data of Ref. \citep{Wanunu_etal_2010} for
$k\approx 7.5$ and $x<5$. Having noted that the calculations presented
in this work are valid to first order in the salt concentration
gradient $(C_t-C_c)/L$, in Fig. \ref{fig:Rate_enhancement_k} we focus
on the regime where $x\leq 5$, as these data points are all for a 1 M
salt solution in the {\it trans} chamber, and represent about 75\% of
the available data. Note in this context that the experimental data of
Ref. \cite{Wanunu_etal_2010} also contain four additional data
points with larger salt concentrations in the {\it trans} chamber (2 M
and 4 M), and larger values of $x$. These data points lie outside the
region where we expect our model to be valid.

To further compare our predictions with the recent experimental
measurements of DNA translocation in salt gradients
\citep{Wanunu_etal_2010}, we fix the system parameters to the
experimental values (for DNA length $N = 400$ bp and $N = 2000$ bp);
$C_t= 1$ ${\rm M}$, $d= 3.5$ ${\rm nm}$, $V= 300$ ${\rm mV}$ and $L =
20$ nm. For the DNA electrophoretic mobility we use $\mu = -10^{-8}
{\rm m^2 s^{-1}V^{-1}}$ \citep{Hartford_Flygare_1975} which for water
with viscosity $\eta = 1$ ${\rm mPa\cdot s}$ (at $T =20^\circ$C) gives
from Eq. (\ref{eq:EP}) $\beta e\phi_{\rm DNA}\approx -0.55$. The only
free parameters are the depletion length $\ell$ and the slip length
$b$.  For depletion lengths $\ell = 0.3$ to $0.6$ nm, which one expects due
to finite-size effects of (hydrated) ions like K$^+$ and Cl$^-$
\cite{Levin_Mena_2001} and image charge effects
\cite{Hatlo_etal_2008}, one finds from $k=7.5$ that $b=4$ to $7$ nm (see
SOM for details), in reasonable agreement with measured values of the
slip length at smooth surfaces \cite{Bonaccurso_etal_2003}. In the
Supplementary Online Material we plot combinations of $\ell$ and $b$
corresponding to different values of $k$, as well as the osmotic flow
profile $v_z(\rho)$ (see Sec. II).

The ions are assumed to be depleted from the pore walls, an assumtion
which is based on numerous theoretical \cite{Levin_etal_2009,
Hatlo_etal_2008,Levin_Mena_2001}, simulation
\cite{Zhu_etal_2005,Chu_etal_2009,Schwierz_etal_2010,Marrink_Marcelja_2001,Huang_etal_2008}
and experimental studies
\cite{Weissenborn_Pugh_1996,Padmanabhan_etal_2007}, that find
nonpolarizable ions to be repelled by an interface between water and a
low dielectric material such as Silicon Nitride ($\epsilon=7$). This
behavior can be modified due to surface chemistry, such as dangling
atoms \cite{Chu_etal_2009}, surface charge and affinity for water
\cite{Schwierz_etal_2010}. Dangling atoms lead to binding of ions to
the surface, which can result in current rectification
\cite{Chu_etal_2009}. Unless these effects are very strong, the ions
are generally all over depleted from neutral low dielectric interfaces.

To conclude, having approximated the ion-wall interactions due to
image charges and water structure by an effective depletion length
$\ell$, we show that the experimental data of Wanunu \textit{et al.}
for DNA translocation in salt gradients can be explained by a
combination of electrophoretic effects and osmosis.  To account for
the different behavior of water on the nanoscale we have also
introduced hydrodynamic slip at the pore walls, which enhances the
flow due to osmosis. With resonable values for both the slip length
and the ion depletion length, we find quantitative agreement between
theory and experimental measurements.

Throughout the calculation we have focused on the diffusion limited
regime, and do not take into account the free-energy barrier felt by
the polymers when entering the pore, yet we are able to quantitatively
reproduce the relative capture rate enhancement data in the
barrier-limited regime. This is most likely a signature of the fact
that the barrier height is nearly constant as a function of salt
asymmetry. We do expect that the main physics reported here, namely
the role of the osmotic flow, explains the enhanced relative
capture rate in the barrier limited regime; however, including the
barrier in our analysis remains a significant challenge.

We have also assumed the ion density inside the accessible part of the
pore to be equal to the average of the salt concentration in the two
reservoirs. This assumption is supported by the current-voltage
relations measured by Wanunu \textit{et al.} for different salt
concentrations in the cis chamber (for $C_t = 1 {\rm M}, C_c = 0.2
{\rm M}$ to $1 {\rm M}$), see supplementary information of Ref.
\citep{Wanunu_etal_2010}.  The calculations presented in this work are
valid to first order in the salt concentration gradient $(C_t-C_c)/L$,
and the results of the measurements by Wanunu \textit{et al.} with
higher salt concentration in the trans reservoir ($x>5$), is outside
the region where this model is expected to be valid.

Putting things in perspective, translocation of DNA through nanopores
is a complicated problem due to its many aspects, ranging from
properties of water in confinement to complicated structures of the
translocating molecules and their interactions.  However, to
understand the recently found increase in capture rate as a function
of salt concentration asymmetry, it seems that a detailed description
of DNA molecules is not needed, since the main mechanism is the
enhanced attraction of DNA molecules towards the pore.  This
attraction is here shown to be made up of two main contributions:
electrophoretic effects and osmotic flow.  The capture rate with
salt gradients due to electrophoretic effects
\citep{Wanunu_etal_2010,Grosberg_Rabin_2010}  and combined with
electro-osmosis \cite{Chou_2009} has been described before, however
the role of (diffusio) osmosis has not been previously discussed. To
understand the osmotic flow it is crucial to account for the repulsive
interaction between kosmotropic ions and a neutral nonpolar wall. We
expect that the main physics is captured by introducing a layer near
the pore wall depleted of ions. This also means that the osmotic flow
is ion specific, and will be reversed when using salt particles that
are net attracted to the pore wall. Finally, our analysis shows that
osmosis cannot be ignored for nanopores in the presence of salt
gradients, even though salt is able to flow through the pore.

\begin{acknowledgments} We thank Meni Wanunu for providing us with
  additional information about the experimental data, and Gerard
  Barkema for useful discussions. This work is part of
  the research program of the
  "Stichting voor Fundamenteel Onderzoek der Materie (FOM)", which is
  financially supported by the "Nederlandse organisatie voor
  Wetenschappelijk Onderzoek (NWO)". \end{acknowledgments}



\newpage
\begin{center}
{\bf SUPPLEMENTARY ONLINE MATERIAL}
\end{center}

In the main text we discuss the dominant contributions to enhanced DNA
translocation induced by a salt gradient in the recent experiment by
Wanunu \textit{et al.}  \cite{Wanunu_etal_20101}. In this
Supplementary Online Material we also discuss other relevant dynamical
processes that can drive DNA towards a nanopore in the presence of
both an external electric field and a salt concentration gradient
accross the pore. We also show plots of the induced osmotic flow
inside the pore, and relevant combinations of the ion depletion length
$\ell$ and the slip length $b$ (see main text for details).

\section{I. RELEVANT DYNAMICAL PHENOMENA FOR DNA TRANSLOCATION IN THE
  PRESENCE OF SALT GRADIENT AND A DRIVING VOLTAGE}

For a narrow pore in the presence of an external electric field, there are two
dynamical phenomena that can drive DNA towards/away from the pore:
\begin{itemize}
\item[(i)] electroosmosis, and
\item[(ii)] electrophoresis.
\end{itemize}
These two mechanisms can be put under the broader group of
electrokinetic effects, which describe the motion of fluids and
charged solutes under forces of electrostatic origin. 
Phenomenon (i) is the movement of a fluid medium containing a net mobile electric charge 
driven by an external electric force; for global neutrality reasons this requires the presence of fixed 
charged walls, as the screening cloud in the fluid constitutes the net mobile charge that couples 
the electric field to the fluid motion.
Phenomenon (ii) concerns the motion of a (often,
confined) charged polymer, such as DNA or RNA, under an applied
electric field, e.g., gel electrophoresis, used for segregating DNA by
their lengths; it describes the motion of a DNA molecule through the pores of a
gel under an external electric field.

In the presence of a salt gradient two further dynamical mechanisms
play a role for the DNA capture by the pore:
\begin{itemize}
\item[(iii)] osmosis, and 
\item[(iv)] diffusiophoresis.  
\end{itemize}
Osmosis (iii) is the flow of water from a low osmotic pressure to the high
osmotic pressure side, i.e., from the {\it cis\/} to the {\it trans\/}
in the experiments of Wanunu {\it et al}. Osmosis through membranes
that are permeable to ions is often called capillary-osmosis or
diffusio-osmosis \cite{Ajdari_Bocquet_20061}, to distinguish the
mechanism from classical osmosis through a semipermeable membrane.
Similarly, associated with the concentration gradient of solutes is
diffusiophoresis (iv), which describes the movement of macromolecules in a
solute concentration gradient \cite{Palacci_etal_20101}. For both
diffusiophoresis and (diffusio) osmosis, the ions (solutes) must
either be attracted to or repelled by the wall/macromolecule. This
repulsion/attraction leads to a lateral pressure gradient in the
region where the ions interact with the wall/macromolecule, driving
fluid flow or movement of the macromolecule. In the main text we
discuss osmosis (iii) and electrophoresis (ii), which we find to be the dominant
contributions to the enhanced DNA capture rate for the particular
system studied by Wanunu {\it et al.}  \cite{Wanunu_etal_20101} i.e.,
electroosmosis is subdominant. Below we argue that diffusiophoresis (iv)
can also be neglected in relation to electrophoresis.
\begin{figure*}
  \begin{minipage}{0.48\linewidth} \resizebox{\columnwidth}{!}{
      \includegraphics[clip]{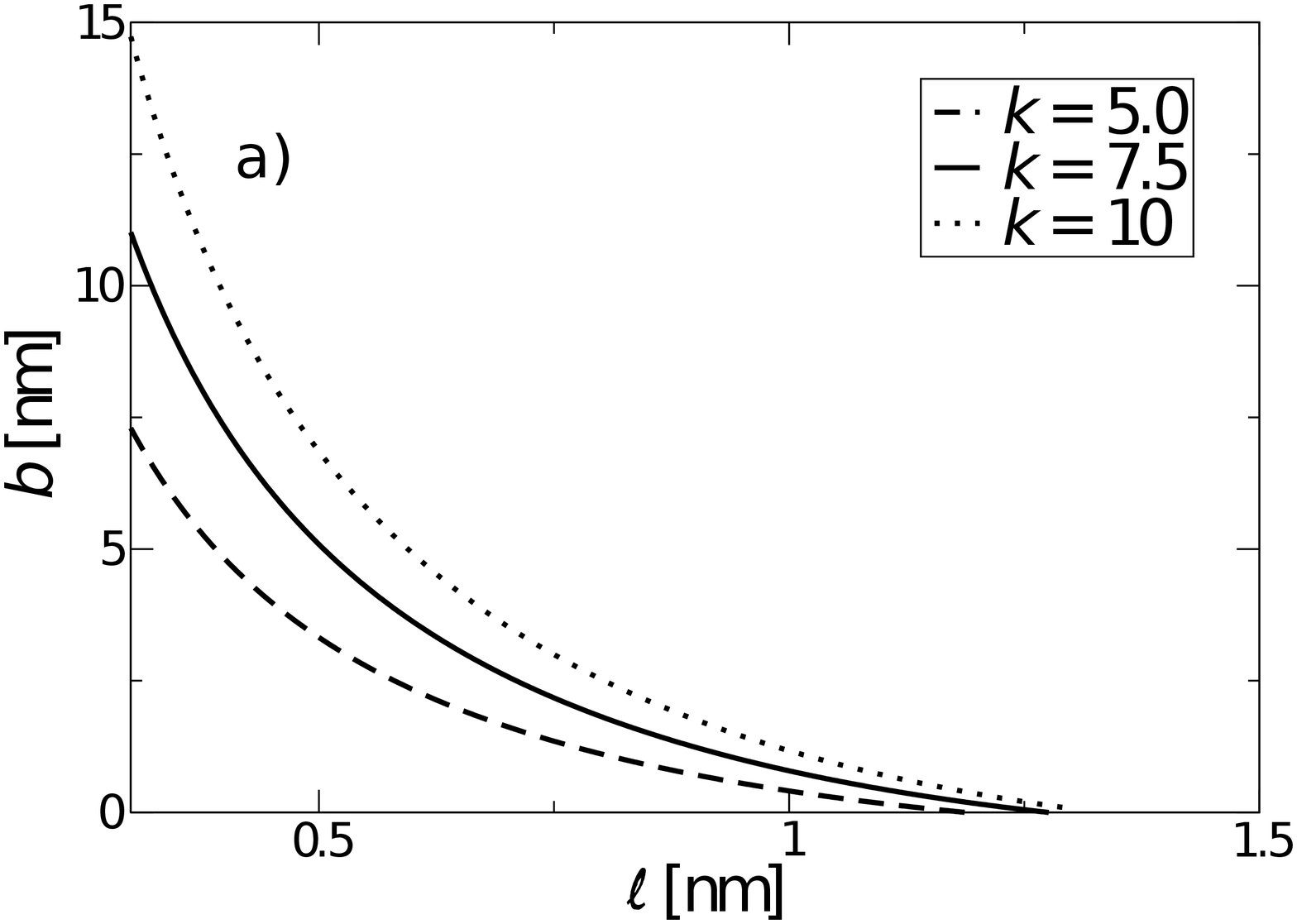}} \end{minipage}
\begin{minipage}{0.48\linewidth} \resizebox{\columnwidth}{!}{
    \includegraphics[clip]{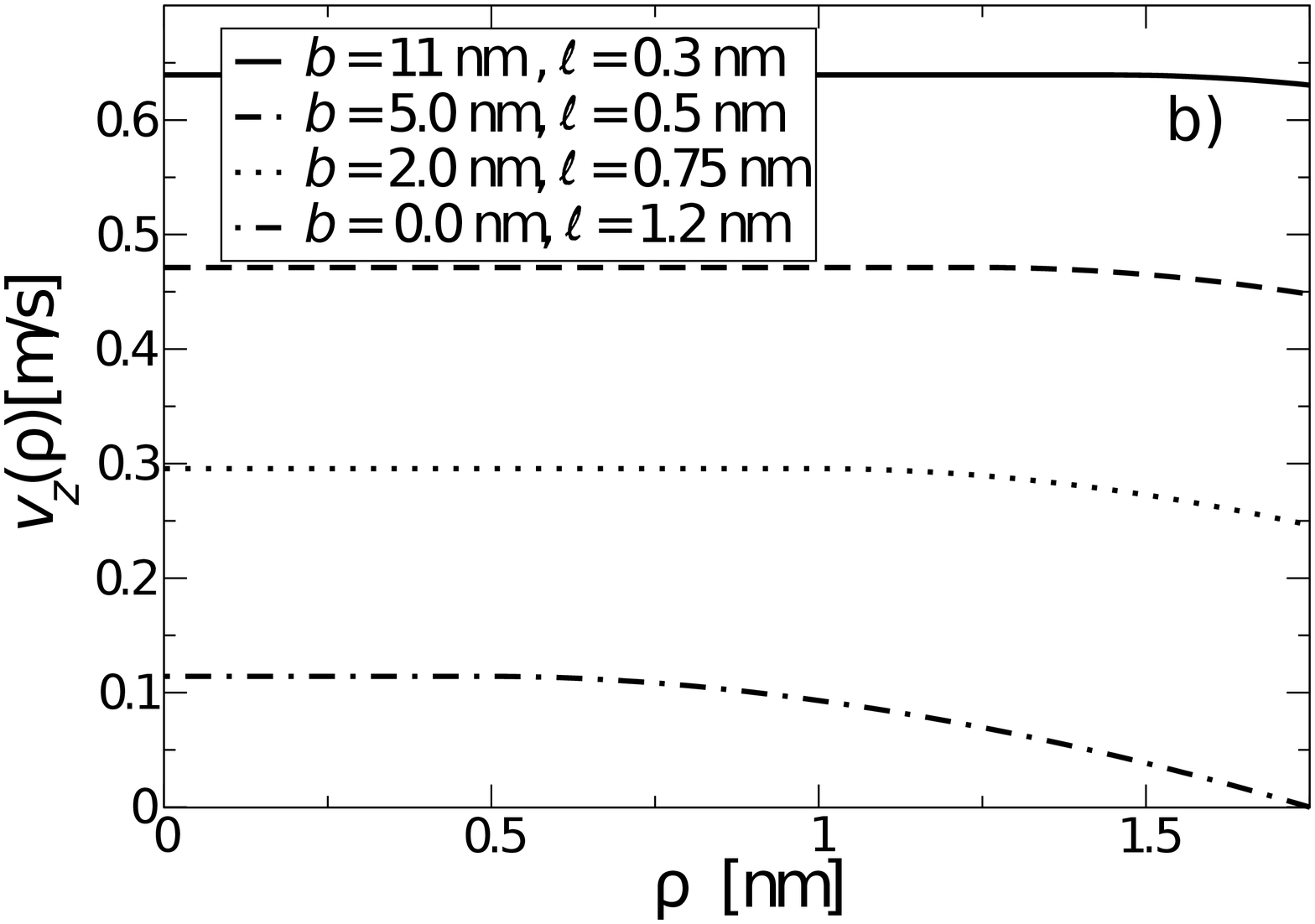}} 
\end{minipage}
\caption{\label{fig:flow}
    (a) Combinations of the slip length $b$ and ion wall depletion
    length $\ell$ for different values of $k$ (see Eq. (17) in the main text),
    with all other values fixed to match the experimental values of
    Ref. \citep{Wanunu_etal_2010}. (b) The liquid flow profile inside
    the pore as a function of the radial coordinate $\rho$ for
    different combinations of the slip length and depletion length
    with $k = 7.5$ and $x=5$, all other
    parameters are fixed to match the experiments of Ref.
    \citep{Wanunu_etal_2010} (see main text for details).}
\end{figure*}
\begin{figure}[!h] \begin{center} \resizebox{\columnwidth}{!}{
      \includegraphics[clip]{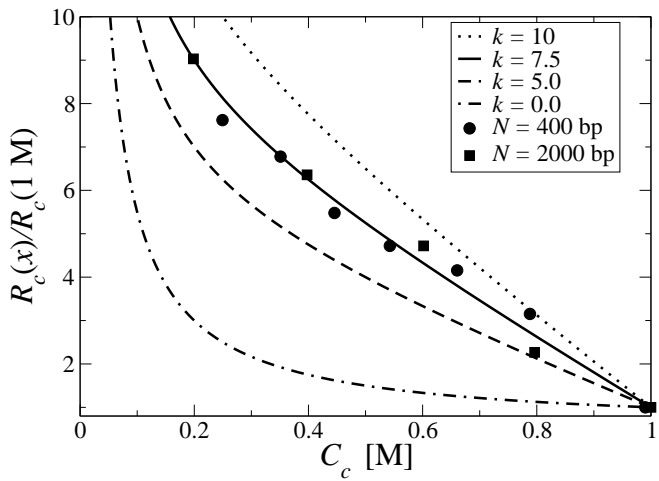}}
    \caption{\label{fig:Rate_enhancement_k_2} Capture rate (Eq. (15)
      in main text) as a function of salt concentration in the {\it
        cis} reservoir, with $C_t=1$ M, for different values of the
      dimensionless parameter $k$ (see Eq. (15) in the main text). The
      points are experimental mesurements of
      Ref. \citep{Wanunu_etal_2010}}
\end{center}
\end{figure}

The flow due to electroosmosis as driven by an electric field $E_p$ in
the pore has the exact same form as the flow due to electrophoresis
(with a change of sign) \cite{Firnkes_etal_20101}
\begin{equation}
v_{EO} = -\frac{\phi_{\rm wall}\epsilon E_{\rm p}}{4\pi \eta},
\end{equation}
so when both the polymer and the pore have surface charge (surface
potential) with the same sign, electrophoresis and electroosmosis pull
the polymer in opposite directions. In the experiment by Wanunu et
al. \cite{Wanunu_etal_20101} the electroosmotic flow was found to be
negligible compared to electrophoresis, as the pore walls was found to
be nearly neutral.  The diffusiophoretic flow, which relates to
diffusioosmosis as electrophoresis relates to electroosmosis, is
induced by the salt gradient in the {\it cis} reservoir, and pushes
DNA towards the pore (as ions are net attracted to DNA molecules). By
conservation of particle current
\begin{equation}
{\bf J} = - D \nabla C({\bf r}),
\end{equation}
the gradient of the salt concentration in the {\it cis} reservoir is
\begin{equation}
\nabla C_c(r) = -\hat{r}\frac{(C_t-C_c)d^2}{8L r^2} 
\end{equation}
and the corresponding flow is
\begin{equation}
\label{eq:DP}
{\bf v_{\rm DP}}(r) = D_{\rm DP}\nabla \ln C_c(r) = -D_{\rm DP}\hat{r}\frac{(x-1)}{8L r^2},
\end{equation}
where $D_{\rm DP}$ is the diffusio-phoretic mobility. The
diffusio-phoretic flow has the same dependence on $x$ as
electrophoresis, and will therefore add to the term linear in $x$ in
the relative capture rate (see Eq. (15) in the main text). The ratio
between the electrophoretic flow and the diffusiophoretic flow is
(from Eq. \eqref{eq:DP} and Eqs. (12) and (13) in the main text)
\begin{equation}
\frac{|{\bf v_{\rm DP}}(r)|}{|{\bf v_{\rm EP}}(r)|} = -\frac{2D_{\rm DP}(x-1)d^2}{\mu V a^2x}\approx 0.1
\end{equation}
with the electrophoretic mobility of DNA $\mu =
-10^{-8}$m$^2$s$^{-1}$V$^{-1}$, $V=300$ mV and $D_{\rm DP} \approx 2
\cdot 10^{-10}$ m$^2$/s \cite{Palacci_etal_20101}. This is about one
order of magnitude lower than electrophoresis and is therefore omitted
in the analysis in the main text.

\section{II. Osmotic flow profile}

In Fig. \ref{fig:flow}(a) we plot the different combinations of $\ell$
and $b$ for different values of $k$, showing resonable combinations of
$\ell$ and $b$ for the $k$-regime of interest (see main text for
details). The slip length of surfaces are typically in the range
$0-30$ nm depending among others on the contact angle and the
smoothness of the surface \citep{Bocquet_Charlaix_20101}.  Similarly,
in Fig. \ref{fig:flow}(b) the liquid flow profile inside the pore is
plotted for different combinations of the depletion length $\ell$ and
the slip length $b$, with $k = 7.5$, $x = 5$, and all other parameters
fixed to match the experiment of Ref.  \citep{Wanunu_etal_2010}. When
the slip length is larger than the pore diameter, $b > d$, the flow
profile becomes almost uniform, in contrast to the no slip case where
the flow is zero near the pore walls. When $\ell$ is small compared to
the pore diameter a larger slip length is needed to produce a large
enough flow to explain the experimental measurements. This is because
a large $\ell$ also decreases the electrophoretic motion of the DNA
molecules, by lowering the ion-acessible area of the pore. For the
flow profiles in Fig.  \ref{fig:flow}, the Reynolds number is $Re =
vd/\nu < 0.0035$, where $\nu = 10^{-6} {\rm m^2/s}$ is the kinematic
viscosity of bulk water.


In Fig. \ref{fig:Rate_enhancement_k_2} we plot the relative capture
rate as a function $C_c$, showing a linear dependence between the
capture rate and $C_c$. This is the exact same plot as Figure 2 in the
main text, however plottet in a different manner. This linear
dependence stems from the osmotic flow as we predict (see Eq. (15) in
the main text), and not from electrophoresis, suggesting that the
capture rate has a linear dependence on $C_c$ for weak salt gradients,
and an inverse dependence for large salt gradients ($x=C_t/C_c$).

\end{document}